# A Data Envelopment Analysis Approach to Benchmark the Performance of Mutual Funds in India

Adit Chopra

Shaheed Sukhdev College of Business Studies, University of Delhi

## Abstract

As the Indian economy grows digitally and becomes more financially inclusive, more and more investors have started to invest in the Indian capital markets. The number of retail and institutional folios with Indian mutual fund schemes have continued to rise for the 74th consecutive month. This study, considers 139 mutual fund schemes (98 equity schemes) and aims to ascertain the various metrics and parameters, retail and institutional investors continue to rely on to make investment recommendations. We compare these with the results from a data envelopment analysis model that generates an efficiency frontier based on an optimal risk, cost and return trade-off. We further put forth an iteration of the DEA model, not only considering risk, cost and return characteristics but also incorporating metrics such as the information ratio which hold significance for retail and institutional investors. We compare these results with traditional metrics and fund rankings published by established industry rating agencies.

## Section1: Introduction

The Indian mutual fund industry is one of the fastest growing and competitive segments of the financial sector. The AUM for the Indian MF industry has grown 4-fold between 2010 and 2020 from Rs. 6.3 trillion to Rs. 25.49 trillion. The number of folios of retail and institutionalized investors rise every day as the Indian financial market becomes more digitally connected and financially inclusive. With these rapid changes in the Indian financial landscape, there is a growing need to ascertain the factors at play when coming up with investment recommendations. It is also worthwhile to understand how retail investors make investment decisions and how we can develop a more comprehensive and holistic framework to compare Indian mutual funds with each other to generate smarter investment recommendations.

Sharpe (1966) put forth a metric that has gained some prominence when evaluating mutual fund performances, the Sharpe ratio which measures the reward-risk trade-off for a mutual fund scheme. Ideally a fund with a high Sharpe ratio is expected to generate higher return per unit and thus funds with higher Sharpe ratios are deemed better than funds with lower Sharpe ratios. This gave rise to a number of traditional metrics being used in the industry by both retail and institutional investors compare funds and make investment recommendations. In section 3, we discuss these traditional metrics and their drawbacks in depth. Over the past few years, investors have started to rely on not only these traditional metrics but also investment recommendations made by established rating agencies in the industry. These rating agencies consider a number of factors to quantify the efficiency of a fund and allot a previously decided weight to each of these factors for each fund to come to a final ranking of funds. However, these metrics themselves do not consider a number of factors such as the cost associated with investment as well as the changes in peer group constituents.

In order to overcome these drawbacks, researchers in recent times have relied on a more statistical methodology of benchmarking mutual fund performances. An approach that can take into account the trade-off between multiple desirable (return) and undesirable factors (risk and cost). One such methodology is the data envelopment analysis model.

Charnes, Cooper and Rhodes (1978) initially proposed an optimization model called data envelopment analysis (DEA) to benchmark the efficiency of public sector activities and non-profits against one another to provide a much better framework of relative efficiencies. DEA is inherently a fractional LPP that is converted into an equivalent LPP to solve for the efficiency measures. Apart from its initially purpose (non-profits and public sector activities), multiple for-profit organizations have utilized the DEA methodology to tabulate efficiencies of various decision-making units (See Seiford (1966))

This paper aims at examining the use of the DEA methodology in order to create a wholesome mutual fund benchmarking index that takes into account multiple input and output factors to calculate the relative efficiencies of each mutual fund (henceforth DMU). Unlike traditional mutual fund performance parameters such as the Sharpe ratio, Treynor's measure, information ratio, Jensen's alpha, etc, we also consider factors such as investment costs, management fees, redemption costs (depicted by proxies such as expense ratio and exit load) in addition to risk measures (beta, standard deviation, downside probability and VaR as a % of corpus) to provide a more comprehensive and detailed efficiency score.

Further, the DEA model presents decision making units as points on or around an efficiency frontier (with efficient units lying on the frontier). As a result, we can utilize the DEA model to calculate the virtual efficiency point for each inefficient DMU giving more insights into the relative performance of each fund as compared to conventional MF ratios.

## Section 2: The data envelopment analysis approach explained

Originally, Charnes, Cooper and Rhodes (1978) introduced a measure of efficiency, defined as the ratio of weighted output to weighted inputs. The concept of weighted sums was introduced to compress the multiple input-output scenario into a single virtual input-virtual output scenario. This proves to be incredibly useful in benchmarking mutual fund returns as comparing individual metrics defining risk, cost and return for a large number of funds is cumbersome. By standardizing the combination of different factors into one factor, that can be used to give an output within a specified range increases our ability to benchmark relative fund performance and generate investment recommendations. The efficiency score is defined as the ratio between the virtual output and the virtual input. The higher the efficiency score, the more efficient a decision-making unit is.

One key determinant of this model is the selection of the weights. Since this was essentially an optimization problem, setting the weights objectively beforehand didn't make sense. Since the DEA model was postulated to measure a trade-off between risk and return characteristics, each risk return combination should be compared to the remaining risk return combinations in the peer set and then standardize the return/risk ratio to a scale of 0 to 1 for each decision making unit by allotting certain weights to each return and risk characteristic.

CCR stated that the efficiency measure is to be calculated by assigning the most favourable weight to each DMU. This implies, that the weights would necessarily not be the same for different DMUs. Thus, the selection of the most favourable weights was done in the following way. Each decision-making unit had the most optimal combination of weights to maximize its efficiency as long as the efficiency of other DMUs in the peer set, computed using the same weighs remained between 0 and 1. Thus, funds with efficiency scores of 1 were considered the most efficient DMUs.

DEA efficiency measure for the decision-making unit $j_0 = 1,2, \ldots n$ is computed by solving the following fractional linear programming model:

$$\max_{\{v_i, u_r\}} \quad h_0 = \frac{\sum_{r=1}^{t} u_r y_{rj_0}}{\sum_{i=1}^{m} v_i x_{ij_0}}$$

Subject to:

$$\frac{\sum_{r=1}^{t} u_r y_{rj}}{\sum_{i=1}^{m} v_i x_{ij}} \leq 1 \qquad j = 1, \ldots, n$$

$$u_r \geq \epsilon \qquad r = 1, \ldots, t$$

$$v_i \geq \epsilon \qquad i = 1, \ldots, m$$

The above ratio has infinite number of optimal solutions. Charnes and Cooper (1962) proposes to select a representative solution from each equivalence class. The representative solution that is usually chosen in DEA modelling is that for which $\sum_{i=1}^{m} v_i x_{ij_0} = 1$ in the input-oriented forms and that with $\sum_{r=1}^{t} u_r y_{ij_0} = 1$ in the output-oriented models. This is how we can convert the fractional LPP into an equivalent LPP to generate the CCR model which generate CRS results (constant returns to scale).

However, a convexity constraint can be appended to the above model that'll make the model correspond to the BCC formulation of DEA which generates VRS (variable returns to scale). CRS reflects the fact that output will change by the same proportion as inputs are changed (e.g. a doubling of all inputs will double output); VRS reflects the fact that production technology may exhibit increasing, constant and decreasing returns to scale. In this study, since input and output variables are not that directly correlated, we can consider using the VRS model.

### Section 3: Traditional metrics for computing mutual fund performances

Both retail and institutional investors frequently rely on traditional metrics to judge fund performance and generate investment recommendations. These metrics are almost always static in nature i.e they are intrinsic to a particular fund. The values of these metrics are not dynamic, in the sense, that they do not vary when the set of funds, a particular fund is being compared with, changes. Thus, they do not provide us with enough information to infer *relative* fund performances. Certain metrics, which are commonly relied upon by investors have been proposed in literature and are discussed in this section.

Starting with the most preliminary measures of fund performance, Beta. This simply is a measure of systematic risk. It is a measure of the fund's volatility compared with its benchmark. Similarly, the Jensen's alpha measures the average return of the portfolio measured over and above that was predicted by CAPM given the portfolio's beta and the average market return.

One of the most used performance indicators is the reward to variability ratio, usually referred to as the Sharpe Ratio, proposed by Sharpe (1966). This is the ratio of the average portfolio excess return over a sample period to the standard deviation of returns over that period. It measures the reward to total volatility trade-off:

$$SR_P = \frac{r_p - r_f}{\sigma_p}$$

The Sharpe ratio considers the excess portfolio return over the risk-free rate per unit of risk, here measured by the standard deviation. However, many investment advisors believe this measure doesn't fully encapsulate the reward to risk trade-off. For one reason, this measure considers reward to be the excess return over the risk-free rate. However, for investors desiring a minimum acceptable return from a particular investment, the return above the risk-free rate isn't a suitable measure. Rather, the return over and above a minimum acceptable return desired by an investor is a better measure of reward. Further, Sharpe ratio considers the standard deviation of return to be the measure of risk borne by an investor. However, in reality the actual measure of risk is the downside movement of the value of investment. Thus, using downside deviation, rather than a generic standard deviation value is more desirable to optimally measure the risk. This measure of fund performance is known as the Sortino ratio. It is the ratio between the excess return over a minimum acceptable return over a time period and the downside deviation of these returns over the same time period.

$$SoR_p = \frac{r_p - MAR}{DD}$$

$$DD = \sqrt{\frac{1}{n} \sum_{i:r_i<MAR} (r_i - MAR)^2}$$

Sortino ratio measures the mutual fund performance against the downside deviation. This ratio is similar to Sharpe ratio, except it takes into account a minimum acceptable return rather than the risk-free rate.

Multiple other metrics have been considered for evaluation of mutual fund performance over time. One of the most commonly used metrics by retail and institutionalized investors is the information ratio. The information ratio is considered to be a benefit/cost ratio for an actively managed fund. The information ratio (IR) is a measurement of portfolio returns beyond the returns of a benchmark, usually an index, compared to the volatility of those returns. The IR is often used as a measure of a portfolio manager's level of skill and ability to generate excess returns relative to a benchmark, but it also attempts to identify the consistency of the performance by incorporating a tracking error, or standard deviation component into the calculation.

The tracking error identifies the level of consistency in which a portfolio "tracks" the performance of an index. A low tracking error means the portfolio is beating the index consistently over time. A high tracking error means that the portfolio returns are more volatile over time and not as consistent in exceeding the benchmark.

$$IR_P = \frac{\alpha_p}{\sigma(\varepsilon_p)}$$

where,

$$\sigma(\varepsilon_p) = \sqrt{\sigma^2(r_p) - \beta_p^2 \sigma^2(r_M)}$$

Other frequently used metrics remain the upside capture ratio and the downside capture ratio. While the upside capture ratio is the measure of the performance of the fund relative to its benchmark during bull markets, the downside capture ratio measures the fund performance relative to its benchmark during a bear market.

*Utility of these conventional mutual fund ratios in decision making:*

These metrics, though conceptually sound and extensive, are difficult to compare and infer from to make sound investment decisions. With so many ratios utilizing different methodologies to measure the risk return trade off, they surely are not a one stop shop to make investment recommendations. Especially when comparing these ratios among 98 different equity mutual funds considered in this study. Although, the viability of some of these ratios can be ascertained by comparing statistically driven results of decision-making units with their respective mutual fund ratios discussed above.

## Section 4: Using a statistical framework to measure risk return trade-offs

In the previous section we discussed about the various traditional frameworks used by both retail and institutionalized investors to evaluate the performance of mutual funds. We explain these measures to some extent but also comment on their static nature and inability to provide clear and concise recommendations when multiple parameters are considered simultaneously, that too for a large sample of funds. Thus, this puts forth a need to have a more comprehensive and relative benchmark to evaluate the efficiencies and performance of these funds from the perspective of an investor. With the large abundance of data points and multiple parameters to be considered, we tend to lean towards establishing a statistical framework that is able to provide us with a definite standard of relative fund performances, benchmarked against one another.

Discussing about statistical frameworks, various techniques have been utilized to evaluate the performances of mutual funds. These techniques usually refer to a multi-criteria decision-making method. This method acknowledges a trade-off between conflicting objectives such as expected return and risk characteristic of a particular mutual fund. One piece of literature elucidating this is the promethee (*preference ranking organization methods for enrichment evaluation*) approach applied on the Italian markets by Cardin, Decima and Pianca (1992)

Our study aims also at optimizing the trade-off between conflicting objectives as well. We wish to generate a framework which takes into account desirable objectives such as expected return (to be maximized) and undesirable objectives such as the risk and cost associated with an investment (to be minimized). Consequently, by allotting weights to each desirable (output factor) and undesirable (input factor), generate an efficiency frontier consisting of those funds with an optimal trade off. Other funds like outside of the efficiency frontier with a virtual efficiency point on the frontier. Thus, we use the multi-criteria decision-making approach in a DEA methodology wherein the weights for each of these factors are not fixed in advance but are optimized for each decision-making unit (not necessarily the same for all decision-making units). Joro, Korhonen and Wallenius (1998) put forth the comparison between the DEA methodology and the multi-criteria decision-making approach extensively.

## Section 5: Variables

Now that we have established the functioning and vision of our statistical framework, we now discuss the various desirable and undesirable factors in our analogy of the trade-off that shall be maximized and minimized respectively.

In our analysis, we have considered the desirable objective to be the expected portfolio return of each decision-making unit that we assign to the output set. Similarly, each undesirable objective (that needs to be minimized) has been assigned to the input set. We have considered the input variables to be broadly of two types: risk defining and cost defining. Each factor that in some way or the other depicts the risk characteristic of the portfolio has been considered as an input factor. Some of these factors we consider in various iterations of our analysis are *standard deviation, beta, downside probability and Value at Risk (as a percentage of corpus).*

The second factor, apart from risk, dictating every investment decision for an investor is the cost associated with it, i.e the subscription and redemption cost. In order to account for this, we have considered the *expense ratio* and the *exit load* of each decision-making unit (mutual fund). These factors have also been assigned to the input set of variables.

In order to achieve a more holistic framework, we have considered various input-output scenarios. While the DEA methodology allows us to consider a multiple input – multiple output scenario, we, at this point in time have only considered one output: the expected return of portfolio. Each different scenario includes an additional input variable depicted above.

*Incorporating the information ratio*

In addition to the above framework wherein the input factors depict the risk and the cost associated with an investment, we also consider another framework depicting a different trade off.

In the above sections, we have discussed some of the traditional metrics for computing mutual fund returns. One of those metrics, the *information ratio*, is of great significance as it points towards the portfolio manager's level of skill and ability to generate excess returns relative to a benchmark, but it also attempts to identify the consistency of the performance by incorporating a tracking error. The Sharpe Ratio simply tells an investor how much he or she was compensated for taking risks, while the Information Ratio tells the investor the rewards the fund manager generated by deviating from the benchmark. A high Information Ratio signals a more consistent and better performing fund, which will give a more consistent investor experience. This is important as investors come in at different points of time. Thus, information ratio, which talks about consistency and not just absolute returns gains an advantage over other traditional measures of fund efficiency.

Thus, given the importance that the information ratio holds in evaluating mutual funds and their performance and efficiencies, we also try to incorporate that into our analysis.

Given that the output factor remains expected portfolio returns, we incorporate the information ratio into our input variables. However, as the multi-criteria decision-making process indicates, input variables or undesirable objectives need to be minimized while desirable objectives or output variables need to be maximized. Thus, incorporating the information ratio as an input factor would dictate the model to categorize funds with lower information ratio as more efficient that funds with higher information ratio. This is counterproductive. Thus, we need to standardize the value of the information ratio for each decision-making unit by subtracting each individual information ratio from a larger number (higher than the largest information ratio of the entire sample). Thus, minimizing the value of $(X - IR)$, where X is the large number and IR is the information ratio, would imply the model to identify decision making units with large information ratios (or smaller $(X - IR)$ values) as more efficient.

Thus, in addition to the original trade-off between expected return and risk, cost characteristics; we also consider the information ratio of funds in addition to their risk characteristic and cost component in a trade-off against the expected portfolio return.

**Table 1.** Here we define the different input-output variable combinations that we would be considering throughout the study. Each of these variables are expected to be a proxy for one of two: Risk characteristics of a particular mutual fund or the cost associated with a particular investment decision.

| Scenario | Input | Output |
|---|---|---|
| 3 Inputs | Beta, Downside Probability, Expense Ratio | Expected Return |
| 4 Inputs | Beta, Downside Probability, Expense Ratio and Standard Deviation | Expected Return |
| 5 Inputs | Beta, Downside Probability, Expense Ratio, Standard Deviation and VaR (as a % of corpus) | Expected Return |
| *Information Ratio Scenarios* | | |
| 4 Inputs | IR, Expense Ratio, DP, Standard Deviation | Expected Return |
| 5 Inputs | IR, Expense Ratio, DP, Standard Deviation, VaR | Expected Return |

7## Section 6: Findings

We initially started with **386** different mutual fund schemes across various categories and sub-categories. We consider returns as *2-year 12 month rolling returns* for each mutual fund as of *29th June 2020* for our analysis. Pursuant to this restriction, we eliminate funds with incorporation after June 29 2017. We further consider funds with a *minimum corpus of Rs. 500 crores*. We also remove funds with incomplete data points. This leaves us with a total of *139 mutual fund schemes*, 98 of which are equity mutual funds belonging to either ELSS, Large Cap, Mid Cap, Multi Cap, Large & Mid Cap or small cap.

<u>Test 1: All Funds:</u> Here we benchmark the performance of all 139 mutual funds in our data set relative to one another. We don't segregate these funds into their categories (equity, debt, hybrid, etc) and sub categories (ELSS, Large Cap, etc) at this stage.

**Table 2.** Efficient funds of the DEA performance index, derived from all 139 mutual funds benchmarked against one another, compared with traditional metrics such as Sharpe, Treynor, Sortino, Information ratio and Jensen's alpha. This table only contains the funds efficient in all or either one of the scenarios. View results for the entire 139 funds in the appendix

| Decision Making Unit | Sharpe | Treynor | IR | Sortino | Jenson | 3 Inputs Rank | 3 Inputs Efficiency | 4 Inputs Rank | 4 Inputs Efficiency | 6 Inputs Rank | 6 Inputs Efficiency | IR_4 Inputs Rank | IR_4 Inputs Efficiency | IR 5 Inputs Rank | IR 5 Inputs Efficiency |
|---|---|---|---|---|---|---|---|---|---|---|---|---|---|---|---|
| Axis Long Term Equity Fund - Growth | 0.2 | 2.49 | 0.75 | 0.49 | 3.67 | 13 | 0.8473 | 13 | 0.8473 | 1 | 1.0000 | 13 | 0.8616 | 1 | 1.0000 |
| Axis Midcap Fund - Growth | -0.71 | -7.69 | -1.44 | -0.9 | -2.35 | 14 | 0.8409 | 14 | 0.8409 | 15 | 0.8794 | 1 | 1.0000 | 1 | 1.0000 |
| Canara Robeco Equity Diversified Fund - Growth | -0.41 | -4.55 | 0.37 | -0.6 | 0.66 | 22 | 0.7875 | 23 | 0.7875 | 22 | 0.8165 | 16 | 0.8543 | 1 | 1.0000 |
| Canara Robeco Equity Tax Saver Fund - Growth | 0.12 | 1.4 | 0.62 | 0.23 | 2.02 | 18 | 0.8127 | 18 | 0.8127 | 7 | 1.0000 | 18 | 0.8442 | 1 | 1.0000 |
| Franklin India Feeder - Franklin U.S. Opportunities Fund - Growth | 1.24 | 15.24 | 1.02 | 2.34 | 6.28 | 1 | 1.0000 | 1 | 1.0000 | 1 | 1.0000 | 1 | 1.0000 | 1 | 1.0000 |
| ICICI Prudential Nifty Index Fund - Reg - Growth | -0.05 | -0.57 | 1.65 | -0.08 | 0.43 | 5 | 0.9700 | 5 | 0.9700 | 8 | 0.9700 | 1 | 1.0000 | 1 | 1.0000 |
| Axis Multicap Fund - Reg - Growth | -0.15 | -1.62 | 1.71 | -0.27 | 2.45 | 12 | 0.8547 | 12 | 0.8547 | 14 | 0.8831 | 11 | 0.9091 | 7 | 1.0000 |
| ICICI Prudential Nifty Next 50 Index Fund - Growth | -1.01 | -8.91 | 0.6 | -1.28 | 0.04 | 10 | 0.9109 | 10 | 0.9109 | 12 | 0.9109 | 4 | 1.0000 | 8 | 1.0000 |
| Parag Parikh Long Term Equity Fund - Reg - Growth | -0.66 | -7.32 | -1.35 | -0.87 | -3.36 | 11 | 0.8670 | 11 | 0.8670 | 13 | 0.8849 | 12 | 0.8975 | 9 | 1.0000 |
| SBI Magnum Global Fund - Growth | -0.45 | -4.78 | 0.4 | -0.93 | 0.55 | 25 | 0.7792 | 21 | 0.7977 | 40 | 0.7792 | 17 | 0.8486 | 10 | 1.0000 |
| Axis Bluechip Fund - Growth | -0.66 | -7.39 | -0.43 | -0.84 | -1.92 | 6 | 0.9669 | 6 | 0.9669 | 5 | 1.0000 | 8 | 0.9360 | 11 | 1.0000 |
| ICICI Prudential FMCG - Growth | 0.08 | 0.91 | -0.49 | 0.16 | -1.48 | 26 | 0.7753 | 29 | 0.7753 | 19 | 0.8335 | 19 | 0.8394 | 12 | 1.0000 |
| Tata Ethical Fund - Reg - Growth | -0.61 | -4.82 | 0.07 | -1.1 | -0.15 | 119 | 0.6438 | 117 | 0.6438 | 123 | 0.6438 | 20 | 0.8348 | 13 | 1.0000 |
| HDFC Index Fund-NIFTY 50 Plan | -0.01 | -0.08 | -3.47 | -0.01 | -0.44 | 3 | 1.0000 | 3 | 1.0000 | 4 | 1.0000 | 5 | 1.0000 | 14 | 1.0000 |
| JM Large Cap Fund - Growth | -0.58 | -6.88 | 1.36 | -0.98 | 5.61 | 4 | 1.0000 | 4 | 1.0000 | 6 | 1.0000 | 6 | 1.0000 | 15 | 1.0000 |
| HDFC Index Fund - Sensex Plan | 0.17 | 2.03 | -6.84 | 0.28 | -0.54 | 1 | 1.0000 | 1 | 1.0000 | 1 | 1.0000 | 7 | 1.0000 | 16 | 1.0000 |

The above table indicates the most efficient funds out of the sample size of 139 funds spread across different categories and sub categories. A key point to note is that while traditional metrics might indicate the risk-reward trade-off, generating investment recommendations is definitely tiresome given that different metrics incorporate different aspects to measure the fund performance. However, in the DEA performance index, we have an opportunity to compare different risk, reward metrics at the same time using a multiple input-output model. Further, we are also able to incorporate another important metric that influences investment decisions: cost. By taking into consideration the expense ratio of different funds, alongside their risk characteristics (standard deviation, beta, downside probability, VaR) and their expected returns we have a much more holistic performance index.

In the first 3 scenarios (3 inputs, 4 inputs, 6 inputs), we notice that certain funds having low information ratios are identified as efficient (for e.g.: HDFC index fund – NIFTY 50 Plan and HDFC Index Fund – Sensex Plan) as compared to funds that do have high information ratio (for e.g.: ICICI prudential nifty index fund) but aren't classified as efficient funds. This can be explained by the fact that ICICI prudential nifty index fund, despite having a higher Information ratio has a higher downside probability (0.27 vs 0.24) and expense ratio (0.45 vs 0.30) and a lower expected portfolio return when compared to HDFC Index Fund – Sensex Plan and Nifty 50 Plan. Thus, the information ratio alone cannot be utilized to make investment decisions as it doesn't consider the impact that multiple risk and cost indicators have when benchmarked against return and consequently compared with other sets of risk, cost, and return variables of 138 other funds. Thus, the static nature of the ratio doesn't fully encapsulate the trade-off that an investor makes when investing in a mutual fund scheme.

8Test 2: Segmentation and categorization of equity mutual funds

In this particular iteration, we segment 98 equity mutual funds from the previously considered set of 139 mutual funds. We further categorized these 98 equity mutual funds into their relevant sub-categories, namely, ELSS (tax saver schemes), Large Cap, Large & Mid Cap, Multi Cap, Mid Cap and Small Cap. We then begin the benchmark the performance of funds within a category with its peers. The variables considered in the analysis remain the same as those defined in Table 1. Here, we further enhance the scope of our study and incorporate fund rankings allotted by prominent rating agency, CRISIL. This proves to be a relevant benchmark to compare the performance of the DEA model given how both retail and institutionalized investors depend on CRISIL's framework to dictate/influence investment decisions.

**Table 3.** Results obtained from the DEA model applied on each of the 6 subcategories of equity mutual funds. The last column contains the CRISIL rankings for each mutual fund for the month of June 2020. The variables considered in different iterations are explained in Table 1. The first column corresponds to the different sub-categories of equity mutual funds have been considered. Columns 3 to 7 contain the value of traditional mutual fund ratios. {Best viewed when zoomed}

| Sub Category | Decision Making Unit | Sharpe | Treynor | IR | Sortino | Jenson | 3 Inputs Rank | 3 Inputs Efficiency | 4 Inputs Rank | 4 Inputs Efficiency | 5 Inputs Rank | 5 Inputs Efficiency | IR_4 Inputs Rank | IR_4 Inputs Efficiency | IR_5 Inputs Rank | IR_5 Inputs Efficiency | CRISIL June Rank |
|---|---|---|---|---|---|---|---|---|---|---|---|---|---|---|---|---|---|
| ELSS | Aditya Birla Sun Life Tax Relief 96 | -0.51 | -6.06 | -0.52 | -0.81 | -3.29 | 2 | 1.0000 | 2 | 1.0000 | 3 | 1.0000 | 3 | 1.0000 | 3 | 1.0000 | 2(2) |
| | Axis Long Term Equity Fund | 0.2 | 2.49 | 0.75 | 0.49 | 3.67 | 1 | 1.0000 | 1 | 1.0000 | 1 | 1.0000 | 1 | 1.0000 | 1 | 1.0000 | 2(1) |
| | Canara Robeco Equity Tax Saver Fund | 0.12 | 1.4 | 0.62 | 0.23 | 2.02 | 3 | 1.0000 | 3 | 1.0000 | 2 | 1.0000 | 2 | 1.0000 | 1 | 1.0000 | 1(1) |
| | DSP Tax Saver Fund | -0.24 | -2.92 | 0.16 | -0.43 | 0.63 | 8 | 0.9305 | 8 | 0.9305 | 8 | 0.9305 | 9 | 0.9305 | 9 | 0.9305 | 3(3) |
| | Franklin India Taxshield | -0.55 | -6.04 | -0.7 | -0.67 | -1.28 | 13 | 0.8878 | 13 | 0.8878 | 13 | 0.8878 | 12 | 0.8985 | 12 | 0.8985 | 4(4) |
| | HDFC Taxsaver | -0.81 | -9.23 | -1.15 | -0.93 | -5.63 | 15 | 0.8744 | 15 | 0.8744 | 15 | 0.8744 | 11 | 0.9002 | 11 | 0.9002 | 5(5) |
| | ICICI Prudential Long Term Equity Fund (Tax Saving) - Reg | -0.23 | -2.53 | 0.45 | -0.33 | 1.09 | 16 | 0.8613 | 16 | 0.8613 | 16 | 0.8613 | 7 | 0.9341 | 7 | 0.9341 | 3(3) |
| | IDFC Tax Advantage (ELSS) Fund - Reg | -0.8 | -3.13 | -2.3 | -0.96 | -7.35 | 18 | 0.8107 | 18 | 0.8107 | 18 | 0.8107 | 18 | 0.8518 | 18 | 0.8518 | 4(4) |
| | Invesco India Tax Plan | -0.18 | -2.1 | 0.02 | -0.33 | -0.15 | 10 | 0.9109 | 10 | 0.9109 | 10 | 0.9109 | 6 | 0.9388 | 6 | 0.9388 | 2(3) |
| | Kotak Tax Saver Fund - Reg | -0.18 | -2.13 | 0.33 | -0.32 | 1.31 | 12 | 0.8914 | 12 | 0.8914 | 12 | 0.8914 | 8 | 0.9335 | 8 | 0.9335 | 2(2) |
| | L&T Tax Advantage Fund - Reg | -0.81 | -3.58 | -1.74 | -1.06 | -6.41 | 7 | 0.9508 | 7 | 0.9508 | 7 | 0.9508 | 10 | 0.9279 | 10 | 0.9279 | 3(4) |
| | Mirae Asset Tax Saver Fund - Reg | -0.04 | -0.5 | 0.65 | -0.08 | 1.66 | 9 | 0.9206 | 9 | 0.9206 | 9 | 0.9206 | 4 | 0.9771 | 4 | 0.9771 | 2(3) |
| | Motilal Oswal Long Term Equity Fund - Reg | -0.48 | -6.38 | -0.1 | -0.92 | -2.45 | 11 | 0.9056 | 11 | 0.9056 | 11 | 0.9056 | 17 | 0.8553 | 17 | 0.8553 | 4(3) |
| | Nippon India Tax Saver (ELSS) Fund | -1.31 | -16.86 | -2.5 | -1.27 | -14.67 | 14 | 0.8775 | 14 | 0.8775 | 14 | 0.8775 | 15 | 0.8662 | 15 | 0.8662 | 5(5) |
| | SBI Long Term Equity Fund | -0.76 | -8.65 | -1.9 | -0.96 | -4.28 | 6 | 0.9538 | 6 | 0.9538 | 6 | 0.9538 | 13 | 0.8842 | 13 | 0.8842 | 3(4) |
| | Sundaram Diversified Equity - Reg | -0.77 | -8.85 | -2.15 | -0.96 | -6.81 | 17 | 0.8353 | 17 | 0.8353 | 17 | 0.8353 | 16 | 0.8633 | 16 | 0.8633 | 5(5) |
| | Tata India Tax Savings Fund - Reg | -0.3 | -4.27 | -0.05 | -0.5 | -5.56 | 4 | 0.9820 | 4 | 0.9820 | 4 | 0.9820 | 14 | 0.8696 | 14 | 0.8696 | 3(3) |
| | UTI Long Term Equity Fund (Tax Saving) | -0.46 | -5.33 | 0.1 | -0.75 | -0.44 | 5 | 0.9568 | 5 | 0.9568 | 5 | 0.9568 | 5 | 0.9744 | 5 | 0.9744 | 2(3) |
| Large and Mid Cap | Aditya Birla Sun Life Equity Advantage Fund | -0.75 | -9.06 | -0.62 | -1.09 | -3.58 | 12 | 0.9181 | 12 | 0.9181 | 12 | 0.9181 | 9 | 0.9109 | 9 | 0.9109 | 3(4) |
| | Canara Robeco Emerging Equities | -0.48 | -5.45 | 0.12 | -0.83 | -0.17 | 6 | 0.9737 | 6 | 0.9737 | 6 | 0.9737 | 4 | 1.0000 | 4 | 1.0000 | 2(2) |
| | DSP Equity Opportunities Fund - Reg | -0.42 | -4.95 | -0.3 | -0.71 | 0.28 | 7 | 0.9555 | 7 | 0.9555 | 7 | 0.9555 | 6 | 0.9637 | 6 | 0.9637 | 3(3) |
| | Franklin India Equity Advantage Fund | -0.72 | -7.84 | -0.72 | -0.83 | -1.61 | 16 | 0.8779 | 16 | 0.8779 | 16 | 0.8779 | 12 | 0.8902 | 12 | 0.8902 | 5(5) |
| | HDFC Growth Opportunities Fund | -0.54 | -6.06 | -0.22 | -0.7 | -0.78 | 18 | 0.8568 | 18 | 0.8568 | 18 | 0.8568 | 16 | 0.8333 | 16 | 0.8333 | 4(3) |
| | ICICI Prudential Large & Mid Cap Fund | -0.66 | -7.39 | -0.43 | -0.84 | -1.92 | 11 | 0.9268 | 11 | 0.9268 | 11 | 0.9268 | 13 | 0.8754 | 13 | 0.8754 | 4(4) |
| | IDFC Core Equity Fund - Reg | -0.71 | -7.69 | -1.44 | -0.9 | -2.35 | 15 | 0.8945 | 15 | 0.8945 | 15 | 0.8945 | 8 | 0.9137 | 8 | 0.9137 | 4(4) |
| | Invesco India Growth Opportunities Fund | -0.15 | -1.62 | 1.71 | -0.27 | 2.45 | 3 | 1.0000 | 3 | 1.0000 | 3 | 1.0000 | 1 | 1.0000 | 1 | 1.0000 | 2(2) |
| | Kotak Equity Opportunities Fund - Reg | -0.27 | -3.47 | -0.09 | -0.5 | -0.93 | 2 | 1.0000 | 2 | 1.0000 | 2 | 1.0000 | 11 | 0.8911 | 11 | 0.8911 | 1(2) |
| | L&T Large and Midcap Fund - Reg | -0.85 | -9.85 | -1.48 | -1.22 | -2.97 | 8 | 0.9412 | 8 | 0.9412 | 8 | 0.9412 | 15 | 0.8419 | 15 | 0.8419 | 3(3) |
| | LIC MF Large & Mid Cap Fund - Reg | -0.31 | -3.69 | 0.14 | -0.63 | 1.34 | 10 | 0.9302 | 10 | 0.9302 | 10 | 0.9302 | 17 | 0.8082 | 17 | 0.7951 | 3(1) |
| | Mirae Asset Emerging Bluechip Fund | -0.07 | -0.86 | 0.82 | -0.14 | 3.88 | 1 | 1.0000 | 1 | 1.0000 | 1 | 1.0000 | 3 | 1.0000 | 3 | 1.0000 | 1(1) |
| | Nippon India Vision | -0.93 | -11.97 | -1.86 | -1.14 | -6.09 | 14 | 0.9098 | 14 | 0.9098 | 14 | 0.9098 | 14 | 0.8715 | 14 | 0.8715 | 5(4) |
| | Principal Emerging Bluechip Fund | -0.68 | -7.95 | -0.15 | -1.1 | -1.42 | 9 | 0.9402 | 9 | 0.9402 | 9 | 0.9402 | 7 | 0.9146 | 7 | 0.9146 | 3(3) |
| | SBI Large & Midcap Fund | -0.41 | -4.55 | 0.37 | -0.6 | 0.66 | 13 | 0.9157 | 13 | 0.9157 | 13 | 0.9157 | 10 | 0.8911 | 10 | 0.8911 | 3(3) |
| | Sundaram Large and Mid Cap Fund - Reg | -0.16 | -1.69 | 1.73 | -0.26 | 5.15 | 5 | 1.0000 | 5 | 1.0000 | 5 | 1.0000 | 2 | 1.0000 | 2 | 1.0000 | 3(3) |
| | Tata Large & Mid Cap Fund - Reg | -0.19 | -2.4 | 0.56 | -0.36 | 1.3 | 4 | 1.0000 | 4 | 1.0000 | 4 | 1.0000 | 5 | 0.9756 | 5 | 0.9756 | 2(3) |
| | UTI Core Equity Fund | -0.95 | -10.24 | -1.61 | -1.05 | -3.63 | 17 | 0.8602 | 17 | 0.8602 | 17 | 0.8602 | 18 | 0.7697 | 18 | 0.7697 | 4(5) |
| Large Cap | Aditya Birla Sun Life Frontline Equity Fund - Reg | -0.49 | -6.03 | -1.23 | -0.66 | -5.32 | 5 | 0.9527 | 5 | 0.9527 | 5 | 0.9527 | 6 | 0.9670 | 7 | 0.9670 | 4(4) |
| | Axis Bluechip Fund | 0.57 | 7.63 | 0.96 | 1.32 | 5.35 | 1 | 1.0000 | 1 | 1.0000 | 1 | 1.0000 | 1 | 1.0000 | 1 | 1.0000 | 1(1) |
| | BNP Paribas Large Cap Fund | 0.01 | 0.1 | 0.14 | 0.02 | 0.71 | 11 | 0.8802 | 11 | 0.8802 | 11 | 0.8802 | 4 | 0.8916 | 4 | 1.0000 | 2(2) |
| | DSP Top 100 Equity Fund - Reg | -0.28 | -3.53 | -0.4 | -0.47 | -2.27 | 13 | 0.8111 | 13 | 0.8111 | 13 | 0.8111 | 13 | 0.8206 | 13 | 0.8206 | 4(4) |
| | Franklin India Bluechip | -0.62 | -6.97 | -2.1 | -0.79 | -4.23 | 10 | 0.8980 | 10 | 0.8980 | 10 | 0.8980 | 7 | 0.9387 | 8 | 0.9387 | 4(5) |
| | HDFC Top 100 Fund | -0.26 | -3.18 | -0.65 | -0.36 | -2.53 | 8 | 0.9198 | 8 | 0.9198 | 8 | 0.9198 | 9 | 0.9198 | 10 | 0.9198 | 5(5) |
| | HSBC Large Cap Equity Fund | -0.22 | -2.81 | -0.56 | -0.38 | -2.54 | 14 | 0.7922 | 14 | 0.7922 | 14 | 0.7922 | 14 | 0.8075 | 14 | 0.8197 | 2(3) |
| | ICICI Prudential Bluechip Fund | -0.27 | -3.04 | -1.16 | -0.4 | -1.96 | 3 | 1.0000 | 3 | 1.0000 | 3 | 1.0000 | 3 | 1.0000 | 3 | 1.0000 | 3(3) |
| | JM Large Cap Fund | -0.55 | -6.84 | -0.65 | -0.9 | -3.44 | 4 | 1.0000 | 4 | 1.0000 | 4 | 1.0000 | 5 | 1.0000 | 5 | 1.0000 | 1(1) |
| | Kotak Bluechip Fund - Reg | -0.18 | -2.3 | -0.82 | -0.32 | -2.21 | 12 | 0.8369 | 12 | 0.8369 | 12 | 0.8369 | 12 | 0.8499 | 12 | 0.8499 | 2(2) |
| | Mirae Asset Large Cap Fund - Reg | -0.1 | -1.14 | -0.1 | -0.16 | -0.24 | 2 | 1.0000 | 2 | 1.0000 | 2 | 1.0000 | 2 | 1.0000 | 2 | 1.0000 | 3(3) |
| | Nippon India Large Cap Fund | -0.3 | -3.53 | -0.8 | -0.41 | -2.73 | 9 | 0.9019 | 9 | 0.9019 | 9 | 0.9019 | 10 | 0.9019 | 11 | 0.9019 | 5(5) |
| | SBI Bluechip Fund | -0.36 | -4.35 | -0.93 | -0.56 | -2.78 | 7 | 0.9294 | 7 | 0.9294 | 7 | 0.9294 | 8 | 0.9382 | 9 | 0.9382 | 4(4) |
| | Tata Large Cap Fund - Reg | -0.28 | -3.69 | -1.14 | -0.43 | -5.57 | 15 | 0.7549 | 15 | 0.7549 | 15 | 0.7549 | 15 | 0.7601 | 15 | 0.7601 | 4(4) |
| | UTI Mastershare Unit Scheme | -0.19 | -2.18 | 0.24 | -0.31 | 0.25 | 6 | 0.9309 | 6 | 0.9309 | 6 | 0.9309 | 11 | 0.9698 | 6 | 0.9883 | 3(3) |
| Multi Cap | Aditya Birla Sun Life Equity Fund | -0.51 | -5.7 | -0.28 | -0.72 | -1.21 | 7 | 0.9523 | 8 | 0.9523 | 7 | 0.9523 | 6 | 0.9761 | 7 | 0.9761 | 4(3) |
| | Axis Multicap Fund - Reg | 0.39 | 4.73 | 2.07 | 0.94 | 7.87 | 1 | 1.0000 | 2 | 1.0000 | 1 | 1.0000 | 1 | 1.0000 | 1 | 1.0000 | --- |
| | Baroda Multi Cap Fund | -0.7 | -8.49 | -0.71 | -1.02 | -4.01 | 16 | 0.8866 | 16 | 0.8866 | 16 | 0.8866 | 13 | 0.8899 | 13 | 0.8899 | 3(3) |
| | BNP Paribas Multi Cap Fund | -0.51 | -8.34 | 0.55 | -0.87 | -4.95 | 5 | 1.0000 | 5 | 1.0000 | 5 | 1.0000 | 18 | 0.8446 | 18 | 0.8446 | 3(3) |
| | Canara Robeco Equity Diversified Fund | 0.07 | 0.8 | 1.45 | 0.14 | 3.5 | 11 | 0.9245 | 11 | 0.9245 | 11 | 0.9256 | 11 | 0.9038 | 1 | 1.0000 | 1(1) |
| | DSP Equity Fund - Reg | -0.06 | -0.82 | 0.41 | -0.15 | 2.43 | 10 | 0.9343 | 10 | 0.9343 | 10 | 0.9343 | 15 | 0.8729 | 15 | 0.8729 | 2(2) |
| | Franklin India Equity Fund | -0.68 | -7.45 | -1.46 | -0.82 | -2.52 | 8 | 0.9523 | 7 | 0.9523 | 8 | 0.9523 | 5 | 0.9835 | 5 | 0.9835 | 4(5) |
| | HDFC Equity Fund | -0.36 | -4.33 | -0.24 | -0.49 | -0.84 | 17 | 0.8677 | 17 | 0.8677 | 17 | 0.8677 | 16 | 0.8723 | 16 | 0.8723 | 5(5) |
| | ICICI Prudential Multicap Fund | -0.33 | -3.81 | -0.18 | -0.46 | -0.32 | 15 | 0.8945 | 15 | 0.8957 | 15 | 0.8957 | 12 | 0.8957 | 12 | 0.8957 | 4(4) |
| | IDFC Multi Cap Fund - Reg | -0.62 | -7.22 | -0.8 | -0.93 | -3.3 | 9 | 0.9408 | 9 | 0.9408 | 9 | 0.9408 | 4 | 0.9503 | 4 | 0.9503 | 4(3) |
| | Invesco India Multicap Fund - Reg | -0.76 | -9.7 | -0.61 | -1.2 | -4.3 | 14 | 0.9037 | 14 | 0.9037 | 14 | 0.9037 | 14 | 0.8855 | 14 | 0.8855 | 3(3) |
| | Kotak Standard Multicap Fund - Reg | -0.15 | -1.77 | 0.21 | -0.25 | 0.51 | 4 | 1.0000 | 4 | 1.0000 | 4 | 1.0000 | 10 | 0.9420 | 10 | 0.9420 | 3(3) |
| | L&T Equity Fund - Reg | -0.66 | -7.32 | -1.35 | -0.87 | -3.36 | 12 | 0.9116 | 12 | 0.9116 | 12 | 0.9116 | 7 | 0.9760 | 7 | 0.9760 | 4(4) |
| | Motilal Oswal Multicap 35 Fund - Reg | -0.66 | -7.88 | -0.77 | -0.96 | -3.8 | 6 | 0.9884 | 6 | 0.9884 | 6 | 0.9884 | 7 | 0.9760 | 7 | 0.9760 | 4(4) |
| | Nippon India Multi Cap Fund | -0.46 | -5.38 | -0.54 | -0.56 | -2.24 | 19 | 0.8378 | 19 | 0.8378 | 19 | 0.8378 | 19 | 0.8378 | 19 | 0.8378 | 5(5) |
| | Parag Parikh Long Term Equity Fund - Reg | 0.2 | 2.39 | 1.08 | 0.41 | 4.25 | 3 | 1.0000 | 3 | 1.0000 | 3 | 1.0000 | 3 | 1.0000 | 3 | 1.0000 | --- |
| | Principal Multi Cap Growth Fund | -0.78 | -8.65 | -1.19 | -1.01 | -3.42 | 18 | 0.8397 | 18 | 0.8397 | 18 | 0.8397 | 17 | 0.8714 | 17 | 0.8714 | 3(3) |
| | SBI Magnum Multi Cap Fund | -0.32 | -3.73 | -0.02 | -0.52 | -0.21 | 13 | 0.9102 | 13 | 0.9102 | 13 | 0.9102 | 10 | 0.9190 | 11 | 0.9190 | 3(3) |
| | UTI Equity Fund | 0.04 | 0.46 | 1.25 | 0.08 | 4.83 | 2 | 1.0000 | 3 | 1.0000 | 3 | 1.0000 | 2 | 1.0000 | 2 | 1.0000 | 1(1) |
| Small Cap | Aditya Birla Sun Life Small Cap Fund | -1.72 | -23.07 | 0.61 | -1.46 | -0.54 | 10 | 0.9213 | 10 | 0.9213 | 10 | 0.9213 | 10 | 0.8384 | 10 | 0.8384 | 5(4) |
| | Axis Small Cap Fund - Reg | 0.06 | 0.89 | 3.16 | 0.17 | 19 | 1 | 1.0000 | 1 | 1.0000 | 1 | 1.0000 | 1 | 1.0000 | 1 | 1.0000 | 1(1) |
| | DSP Small Cap Fund - Reg | -1.37 | -17.24 | 0.38 | -1.44 | -0.1 | 4 | 0.9839 | 4 | 1.0000 | 4 | 1.0000 | 4 | 1.0000 | 4 | 1.0000 | 3(3) |
| | Franklin India Smaller Companies Fund | -1.4 | -17.63 | 1.21 | -1.29 | 3.19 | 3 | 1.0000 | 3 | 1.0000 | 3 | 1.0000 | 3 | 1.0000 | 3 | 1.0000 | 3(3) |
| | HDFC Small Cap Fund | -0.82 | -14.14 | 1.03 | -0.97 | 7.15 | 5 | 1.0000 | 5 | 1.0000 | 5 | 1.0000 | 5 | 0.9995 | 5 | 0.9995 | 3(3) |
| | ICICI Prudential Smallcap Fund - Ret | -0.8 | -10.51 | 1.27 | -1.1 | 11.06 | 6 | 0.8913 | 6 | 0.8913 | 11 | 0.8913 | 9 | 0.8913 | 9 | 0.8913 | 2(3) |
| | Kotak Small Cap Fund - Reg | -0.9 | -12.54 | 2.64 | -1.28 | 8.45 | 6 | 0.9647 | 6 | 0.9647 | 6 | 0.9647 | 2 | 1.0000 | 2 | 1.0000 | --- |
| | L&T Emerging Businesses Fund - Reg | -1.32 | -17.17 | 0.26 | -1.28 | -0.03 | 6 | 0.9608 | 5 | 0.9716 | 5 | 0.9716 | 6 | 0.9669 | 6 | 0.9669 | 4(3) |
| | Nippon India Small Cap Fund | -1.05 | -12.41 | 1.69 | -1.24 | 4.73 | 8 | 0.9274 | 8 | 0.9274 | 8 | 0.9274 | 7 | 0.9418 | 7 | 0.9418 | 3(3) |
| | SBI Small Cap Fund | -0.56 | -6.7 | 2.58 | -0.95 | 11.45 | 9 | 0.9245 | 9 | 0.9245 | 9 | 0.9245 | 8 | 0.9400 | 8 | 0.9400 | 2(2) |
| | Sundaram Small Cap Fund - Reg | -1.71 | -23.59 | 0.48 | -1.51 | -0.98 | 7 | 0.9425 | 7 | 0.9425 | 7 | 0.9425 | 11 | 0.8291 | 11 | 0.8291 | 4(5) |

| | Fund | | | | | | | | | | | | | | | | | |
|---|---|---|---|---|---|---|---|---|---|---|---|---|---|---|---|---|---|---|
| Mid Cap | Aditya Birla Sun Life Mid Cap Fund - Plan A | -1.39 | -15.13 | -0.59 | -1.36 | -2.12 | 15 | 0.8347 | 15 | 0.8347 | 15 | 0.8347 | 11 | 0.8621 | 11 | 0.8621 | 5(5) |
| | Axis Midcap Fund | 0.34 | 3.49 | 4.8 | 0.88 | 12.58 | 1 | 1.0000 | 1 | 1.0000 | 1 | 1.0000 | 1 | 1.0000 | 1 | 1.0000 | 1(1) |
| | BNP Paribas Mid Cap Fund | -0.89 | -19.31 | -0.77 | -1.26 | -8.75 | 4 | 1.0000 | 4 | 1.0000 | 4 | 1.0000 | 13 | 0.8368 | 13 | 0.8368 | 2(2) |
| | DSP Midcap Fund - Reg | -0.5 | -6.31 | 1.39 | -0.96 | 5.2 | 2 | 1.0000 | 2 | 1.0000 | 2 | 1.0000 | 3 | 1.0000 | 3 | 1.0000 | 2(2) |
| | Edelweiss Mid Cap Fund | -0.78 | -13.41 | -0.85 | -1.2 | -8.09 | 9 | 0.9455 | 9 | 0.9455 | 9 | 0.9455 | 14 | 0.8324 | 14 | 0.8324 | 3(3) |
| | Franklin India Prima Fund | -0.78 | -8.68 | 0.8 | -0.99 | 1.52 | 5 | 0.9822 | 5 | 0.9822 | 5 | 0.9822 | 5 | 0.9825 | 5 | 0.9831 | 4(3) |
| | HDFC Mid-Cap Opportunities Fund | -0.97 | -10.42 | 2.04 | -1.17 | 2.32 | 3 | 1.0000 | 3 | 1.0000 | 3 | 1.0000 | 2 | 1.0000 | 2 | 1.0000 | 3(4) |
| | ICICI Prudential MidCap Fund | -0.99 | -11.1 | -0.37 | -1.11 | -1.53 | 17 | 0.7928 | 17 | 0.7928 | 17 | 0.7928 | 17 | 0.7937 | 17 | 0.7937 | 4(5) |
| | Invesco India Mid Cap Fund | -0.44 | -4.84 | 3.02 | -0.78 | 6.68 | 13 | 0.8843 | 13 | 0.8843 | 12 | 0.8980 | 12 | 0.8511 | 12 | 0.8511 | 1(1) |
| | Kotak Emerging Equity Fund - Reg | -0.58 | -6.88 | 1.36 | -0.98 | 5.61 | 6 | 0.9777 | 6 | 0.9777 | 6 | 0.9777 | 7 | 0.9777 | 7 | 0.9777 | 3(3) |
| | L&T Midcap Fund - Reg | -1.12 | -12.42 | 0.82 | -1.37 | 0.36 | 7 | 0.9617 | 7 | 0.9617 | 7 | 0.9617 | 4 | 0.9932 | 4 | 0.9932 | 3(3) |
| | Motilal Oswal Midcap 30 Fund - Reg | -0.56 | -6.69 | 1.03 | -0.91 | 6.39 | 16 | 0.8345 | 16 | 0.8345 | 16 | 0.8345 | 15 | 0.8298 | 15 | 0.8298 | 4(4) |
| | Nippon India Growth Fund | -0.62 | -6.65 | 1.18 | -0.92 | 4.55 | 11 | 0.9372 | 11 | 0.9372 | 11 | 0.9373 | 9 | 0.9372 | 9 | 0.9373 | 3(3) |
| | SBI Magnum Midcap Fund | -1.27 | -15.03 | -0.64 | -1.39 | -4.55 | 12 | 0.8952 | 12 | 0.8952 | 13 | 0.8952 | 10 | 0.9009 | 10 | 0.9009 | 4(4) |
| | Sundaram Mid Cap Fund - Reg | -1.16 | -12.94 | 0.11 | -1.33 | -0.09 | 10 | 0.9417 | 10 | 0.9417 | 10 | 0.9417 | 8 | 0.9429 | 8 | 0.9429 | 5(4) |
| | Tata Mid Cap Growth Fund - Reg | -0.51 | -6.46 | 1.24 | -0.85 | 5.43 | 14 | 0.8478 | 14 | 0.8478 | 14 | 0.8478 | 16 | 0.8097 | 16 | 0.8097 | 3(2) |
| | UTI Mid Cap Fund | -1.14 | -13.05 | -0.13 | -1.44 | -2.25 | 8 | 0.9554 | 8 | 0.9554 | 8 | 0.9554 | 6 | 0.9787 | 6 | 0.9787 | 3(3) |

**Table 4.** The decision-making units in Table 3 have been colour coded on the basis of certain parameter described as follows

| Key | | |
|---|---|---|
| Colour | Meaning | Number of Funds |
| | Efficient funds in all scenarios (3,4,5 Inputs and all IR scenarios) | 20 |
| | Efficient in IR Scenarios but not in initial 3,4,5 input scenaios | 4 |
| | Efficient in intial 3,4,5 input scenarios but not in IR scenario (s) | 5 |

Table 3 puts forth the DEA efficiency values for each decision-making unit when compared within its sub-category. Table 2 and Table 3 together illustrate one of the key advantages of the DEA methodology over traditional metrics, the ability to be dynamic. The values for the traditional metrics remained the same in both the tables, however, the efficiency and rank of each decision-making unit changed when its peer group was changed. In table 2, we compared each DMU with 139 other DMUs irrespective of the type of mutual fund it was. Equity mutual funds being benchmarked against index funds. ELSS funds being compared with large cap funds. The variable for their sub-category wasn't controlled for. However, in Table 3, each DMU within a sub-category of equity mutual funds is being benchmarked against other DMUs within that same sub-category. Thus, the DEA model adapts to changes in peer group, something the traditional metrics do not.

Looking at the effect of changes in input variable composition on the overall efficiency of each DMU, we see that in the first 3 scenarios (3 inputs, 4 inputs, 5 inputs) the addition of standard deviation in the 2$^{nd}$ iteration and VaR as a % of corpus in the 3$^{rd}$ iteration didn't make a significant difference in the efficiency of each decision making unit. Given, how the 1$^{st}$ iteration already had beta and downside probability as two risk measures, the addition of standard deviation and VaR didn't make a big difference as they themselves are defining the risk characteristic of a mutual fund scheme.

However, we see that incorporating the information ratio as another input variable, after standardizing it, has a significant impact on the efficiency of decision-making units. Table 4 indicates how as many as 4 new mutual fund schemes were identified as efficient after the introduction of the information ratio. Further, 5 schemes, though earlier efficient, were deemed no longer efficient under the scenarios involving the information ratio. These are usually funds with negative or incredibly low values of information ratio.

Similarly, funds such as *Kotak Small Cap – Reg* and *Canara Robeco Equity Diversified Fund* are identified as efficient in the information ratio scenarios considering their relatively high information ratios in their respective peer group. *Kotak Small Cap – Reg*: IR is 2.64, second only to Axis small cap; *Canara Robeco Equity Diversified Fund*: IR is 1.45, second only to Axis Multicap.

### Section 7: Comparative analysis with rating agency frameworks

The last column (column 18) of Table 3 entails the CRISIL ratings for June 2020 for each mutual fund scheme under our consideration. The CRISIL rating framework for equity mutual funds takes into account a variety of factors such as mean return and volatility, active return, portfolio concentration analysis and liquidity analysis. The CRISIL methodology though doesn't employ a relative benchmarking mechanism, it allots a certain fixed weight to each of the above stated parameter to arrive at the final rankings. For example, to calculate the mean return, it assigns progressive weights (32.5%, 27.5%, 22.5% and 17.5%) to each segmented period (latest 36,27,18 and 9 months) of the 3-year period of analysis for equity funds (starting with the longest period) Similarly, they employ certain arithmetic techniques to calculate exposure in each of the other factors. CRISIL ratings are an industry standard to compare mutual fund schemes in India. The DEA model relies more on the core fundamentals of risk, cost and return to benchmark fund performances relative to one another.

**Table 5.** Comparative analysis of DEA model results and CRISIL rankings. The funds have been classified by their CRISIL rankings. Column 2 contains the total number of funds corresponding to the ranking. Column 2 contains the number of these funds that are identified as efficient by the DEA model and Column 3 contains the number of funds identified as inefficient by the DEA model.

| CRISIL Rank | Number of Funds | Number of efficient funds | Number of inefficient funds |
|---|---|---|---|
| 1 | 10 | 9 | 1 |
| 2 | 18 | 10 | 8 |
| 3 | 31 | 7 | 24 |
| 4 | 24 | 1 | 23 |
| 5 | 12 | 0 | 12 |
| **Total** | **95** | **27** | **68** |

Table 5 evaluates the performance of the DEA benchmarking framework with the CRISIL rating framework. Since the CRISIL framework has a limited scope of ranking (between 1 and 5), the comparison between the two frameworks becomes tougher. One key observation that we evidently notice corresponds to the funds ranked 1 according to CRISIL. We can see that out of 10 funds ranked 1, 9 of them have been identified as efficient by the DEA model and 1 fund, *Invesco India Mid Cap Fund* has been identified as inefficient. This can be explained by Invesco India Mid Cap Fund's high expense ratio of 2.35 (among the highest in the peer group), high beta and standard deviation (0.83 and 9.21 respectively) and relatively high downside probability (0.53). Thus, despite having a high information ratio of 3.02 (second highest in the peer group), it is not efficient in either scenario.

## Section 8: Conclusion

In recent times, many initiatives undertaken by the government of India to become more digitally connected and financially inclusive have had a profound impact on the development of the Indian financial markets. The India mutual fund industry has shown signs of incredible growth in recent years. The AUM of the industry have grown from Rs. 6.3 trillion as on June 30, 2010 to Rs. 25.49 trillion as on June 30 2020, a 4-fold increase in a span of 10 years. As the number of folios rise for the 73[rd] consecutive month, to 9.15 crore accounts {8.04 crore of which falling under equity, hybrid and solution-oriented schemes, attracting high investments from the retail segments}, a holistic methodology of evaluating mutual fund schemes and generating investment recommendations continues to gain eminence. This study aims at generating one such methodology for the Indian capital markets, taking into account risk characteristics of funds, costs associated with investment and expected return derived from said investment, to benchmark fund efficiencies relative to one another. Not only this, we also strive to incorporate an important traditional metric, information ratio, into our adaptation of a multi-criteria decision-making model. We consider different input-output combinations to ascertain the utility of variables used to define parameters in our trade-off. Further, we acknowledge how traditional metrics such as Sharpe Ratio, Sortino Ratio, Treynor's Measure and Information Ratio have been of widespread usage both at a retail level and an institutional level, and thus we compare our findings and recommendations with the aforementioned ratios. We conclude that despite the ratios having a sound conceptual backing, they lack to provide concrete investment recommendations by themselves. Further, these ratios don't take into account the costs associated with the investment under their purview. Thus, when multiple factors need to be accounted for in a trade-off, using traditional metrics to make decisions is cumbersome. Further, we increase the scope of our study to compare results from our DEA model with the corresponding fund rankings given by CRISIL, an industry wide recognized rating agency. We observe that only on one 1 such occasion, a CRISIL ranked 1 fund hasn't been considered efficient in either of the iterations of the DEA model. This study is an attempt to understand the factors at play when evaluating fund performances and generating investment recommendations, and consequently put forth a statistical framework to benchmark relative fund efficiencies and compare findings with both established research and traditional metrics in the Indian context.

## Section 9: References

## Section 10: Annexure

**Table 6.** Results in continuation to Table 2. This table contains the efficiency results for all of the 139 mutual funds when benchmarked against each other. We consider 5 different iterations (3 Inputs, 4 Inputs, 5 Inputs and two information ratio scenarios). Zoom in to view results.